\documentclass[twocolumn,showpacs,preprintnumbers,amsmath,amssymb]{revtex4}
\usepackage{graphicx}
\usepackage{dcolumn}
\usepackage{bm}
\begin{document}

\title{The distribution of local fluxes in porous media}

\author{Asc\^anio D. Ara\'ujo}
\author{Wagner B. Bastos}
\author{Jos\'e S. Andrade Jr.}
\author{Hans J. Herrmann}

\altaffiliation[Formerly at ] {Institute for Computer Physics,
University of Stuttgart.}

\affiliation{Departamento de F\'{\i}sica, Universidade Federal 
do Cear\'a,\\ 60451-970 Fortaleza, Cear\'a, Brazil.}

\date{\today}

\begin{abstract}
We study the distributions of channel openings, local fluxes, and
velocities in a two-dimensional random medium of non-overlapping
disks. We present theoretical arguments supported by numerical data of
high precision and find scaling laws as function of the porosity. For
the channel openings we observe a crossover to a highly correlated
regime at small porosities. The distribution of velocities through
these channels scales with the square of the porosity. The fluxes turn
out to be the convolution of velocity and channel width corrected by a
geometrical factor. Furthermore, while the distribution of velocities
follows a Gaussian, the fluxes are distributed according to a
stretched exponential with exponent 1/2. Finally, our scaling analysis
allows to express the tortuosity and pore shape factors from the
Kozeny-Carman equation as direct average properties from microscopic
quantities related to the geometry as well as the flow through the
disordered porous medium.

\end{abstract}


\maketitle

Fluid flow through a porous medium is of importance in many practical
situations ranging from oil recovery to chemical reactors and has been
studied experimentally and theoretically for a long time
\cite{Dullien79,Adler92}. Due to disorder, porous media
display many interesting properties that are however difficult to
handle even numerically. One important feature is the presence of
heterogeneities in the flux intensities due the varying channel
widths. They are crucial to understand stagnation, filtering,
dispersion and tracer diffusion. These are subjects of much practical
interest in medicine, chemical engineering and geology and on which a
vast literature is available \cite{Sahimi95}.

Many stochastic models for disordered porous media have been proposed
and used to describe the above mentioned effects. One of the most
successful is the so-called {\it q-model} for force distributions in
random packings \cite{Coppersmith96} in which a scalar fluid is
transfered downwards from layer to layer. Although the distribution of
local flux intensities should be the basis for any quantitative
evolution of these stochastic models, detailed studies of them at the
pore level are still lacking.

The traditional approach for the investigation of single-phase fluid
flow at low Reynolds number in disordered porous media is to
characterize the system in terms of Darcy's law \cite{Dullien79,Sahimi95}, 
which assumes that a {\it macroscopic} index, the permeability $K$,
relates the average fluid velocity $V$ through the pores with the
pressure drop $\Delta P$ measured across the system,
\begin{equation}
V = -{K \over \mu}{\Delta P \over L}~,
\label{eq1}
\end{equation}
where $L$ is the length of the sample in the flow direction and $\mu$
is the viscosity of the fluid. In fact, the permeability reflects the
complex interplay between porous structure and fluid flow, where {\it
local} aspects of the pore space morphology and the relevant mechanisms 
of momentum transfer should be adequately considered. In previous studies 
\cite{Canceliere90,Kostek92,Martys94,Andrade95,Koponen97,Rojas98,Andrade99}, 
computational simulations based on detailed models of pore geometry
and fluid flow have been used to predict permeability coefficients as
well as to validate semi-empirical correlations obtained from real porous
materials.

In this paper we present numerical calculations for a fluid flowing
through a two-dimensional channel of width $L_y$ and length $L_x$
filled with randomly positioned circular obstacles. For instance, this
type of model has been frequently used to study flow through fibrous
filters \cite{Marshall94}. Here the fluid flows in the $x$-direction at
low but non-zero Reynolds number and in the $y$-direction we impose
periodic boundary conditions. We consider a particular type of random
sequential adsorption (RSA) model \cite{Torquato02} in two dimensions
to describe the geometry of the porous medium. As shown in Fig.~1,
disks of diameter $D$ are placed randomly by first choosing from a
homogeneous distribution between $D/2$ and $L_x-D/2$ ($L_y-D/2$) the
random $x$-($y$-)coordinates of their center. If the disk allocated at
this position is separated by a distance smaller than $D/10$ or
overlaps with an already existing disk, this attempt of placing a disk
is rejected and a new attempt is made. Each successful placing
constitutes a decrease in the porosity (void fraction) $\epsilon$ by
$\pi D^{2}/4 L_x L_y$. One can associate this filling procedure to a
temporal evolution and identify a successful placing of a disk as one
time step. By stopping this procedure when a certain value of
$\epsilon$ is achieved, we can produce in this way systems of well
controlled porosity. We study in particular configurations with
$\epsilon=0.6$, $0.7$, $0.8$ and $0.9$.   
\begin{figure}
\includegraphics[width=8cm]{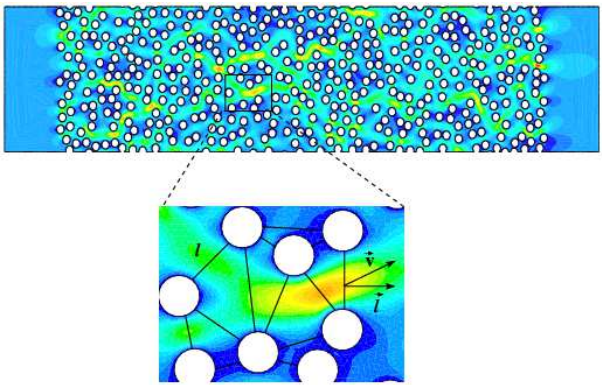}
\caption{Contour plot of the velocity magnitude for a typical 
realization of a pore space with porosity $\epsilon=0.7$ subjected to
a low Reynolds number and periodic boundary conditions applied in the
$y$-direction. The fluid is pushed from left to right. The colors
ranging from blue (dark) to red (light) correspond to low and high
velocity magnitudes, respectively. The close-up shows a typical pore
opening of length $l$ across which the fluid flows with a line average
velocity $\vec{v}$. The local flux at the pore opening is given by
$q=vl\cos\theta$, where $\theta$ is the angle between $\vec{v}$ and the
vector normal to the line connecting the two disks.}
\label{fig1}
\end{figure}

First we analyze the geometry of our random configurations making a
Voronoi construction of the point set given by the centers of the
disks \cite{Voronoi08,Watson81}. We define two disks to be neighbors
of each other if they are connected by a bond of the Voronoi
tessellation.  These bonds constitute therefore the openings or pore
channels through which a fluid can flow when it is pushed through our
porous medium, as can be seen in the close-up of Fig.~1. We measure the
channel widths $l$ as the length of these bonds minus the diameter $D$
and plot in Fig.~2 the (normalized) distributions of the normalized
channel widths $l^* = l/D$ for the four different porosities. Clearly
one notices two distinct regimes: ({\it i}) for large widths $l^*$ the
distribution decays seemingly exponentially with $l^*$, and ({\it ii})
for small $l^*$ it has a strong dependence on the porosity, increasing
dramatically at the origin with decreasing porosity.
\begin{figure}
\includegraphics[width=8cm]{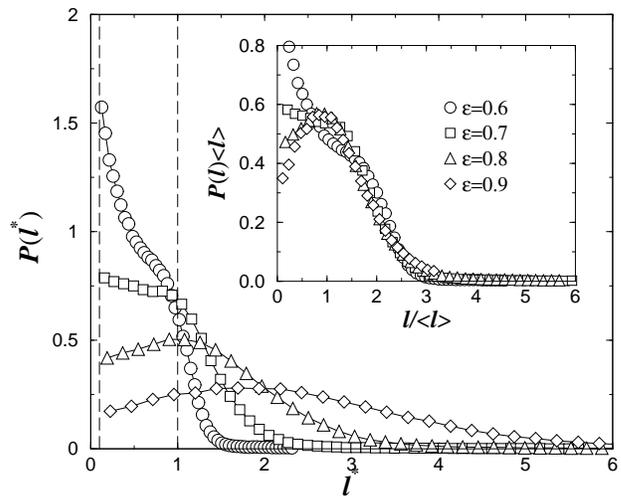}
\caption{Distributions of the normalized channel widths $l^{*}=l/D$
for different values of porosity $\epsilon$. From left to right, the
two vertical dashed lines indicate the values of the minimum distance
between disks $l^{*}=0.1$ and the size of the disks $l^{*}=1$. The
inset shows the data collapse obtained by rescaling the distributions
with $\langle l \rangle$ using Eq.~(\ref{Eq_Hans}).}
\label{fig2} 
\end{figure}
A closer investigation shows that in Fig.~2 the large $l^*$ tail
decays like a Gaussian for large porosities while it is a simple
exponential when the porosity is around or below $0.7$.  The crossover
between the two regimes is visible as a peak which shifts between
$\epsilon = 0.9$ and $0.8$ and then stays for smaller porosities at
about $l^*=1$, i.e., $l = D$. These distribution functions can be
qualitatively understood in the following way. For very large
porosities, i.e., very dilute systems, the distance between the
particles is essentially uncorrelated due to excluded volume and is
therefore Gaussian distributed around a mean value $\langle l \rangle
$. If for simplicity one imagines particles being on a regular
triangular lattice as an idealized configuration in two dimensions,
the following expression is obtained:
\begin{equation}
\langle l \rangle = D(\sqrt{{\pi\over 2 \sqrt{3} (1-\epsilon)}}-1)~.
\label{Eq_Hans}
\end{equation}
The filling process will strongly feel the clogging due to excluded
volume when one disk just fits into the hole between three disks. This
situation occurs when $\langle l \rangle = D (\sqrt{3} -
1)$. Inserting this into Eq.~(\ref{Eq_Hans}) gives a crossover
porosity of $\epsilon=1-\pi/6\sqrt{3} \approx 0.7$ which agrees with
our simulation (see Fig.~2). Interestingly, a related property, namely
the correlation function, does not seem to show such a crossover
\cite{Torquato95,Rintoul96}. The inset of Fig.~2 shows that, for
sufficiently large values of $l$, all distributions $P(l)$ collapse to
a single curve when rescaled by the corresponding value of $\langle l
\rangle$ calculated from Eq.~(\ref{Eq_Hans}). As shown in the inset of 
Fig.~3, the variation of the average value $\langle l^{*} \rangle$
with porosity follows very closely Eq.~(\ref{Eq_Hans}). Only the
prefactor is different from unity ($\approx 1.2$) due to the presence
of disorder. This result indicates that our simple description based
on a diluted system of particles placed on a regular lattice provides
a good approximation for the geometry of the disordered porous medium.

The fluid mechanics in the porous space is based on the assumption
that a Newtonian and incompressible fluid flows under steady-state 
conditions. The Navier-Stokes and continuity equations for this case 
reduce to
\begin{equation}
\rho~{\vec{u}\cdot \nabla \vec{u}} = -{\nabla p} + 
\mu~{\nabla}^{2}{\vec{u}}~,
\label{Eq_momentum}
\end{equation}
\begin{equation}
{\nabla \cdot \vec{u}}=0~,
\label{Eq_continuity}
\end{equation}
where $\vec{u}$ and $p$ are the local velocity and pressure fields,
respectively, and $\rho$ is the density of the fluid. No-slip boundary
conditions are applied along the entire solid-fluid interface, whereas
a uniform velocity profile, $u_{x}(0,y)=V$ and $u_{y}(0,y)=0$, is
imposed at the inlet of the channel. For simplicity, we restrict our
study to the case where the Reynolds number, defined here as
$Re\equiv{\rho V L_{y} / \mu}$, is sufficiently low ($Re < 1$) to
ensure a laminar viscous regime for fluid flow. We use FLUENT
\cite{Fluent}, a computational fluid dynamic solver, to obtain the
numerical solution of Eqs.~(\ref{Eq_momentum}) and
(\ref{Eq_continuity}) on a triangulated grid of up to hundred
thousand points adapted to the geometry of the porous medium.

Simulations have been performed by averaging over $10$ different pore
space realizations generated for each value of porosity. The contour
plot in Fig.~1 of the local velocity magnitude for a typical
realization of the porous medium with porosity $\epsilon=0.7$ clearly
reveals that the transport of momentum through the complex geometry
generates preferential channels \cite{Andrade99}. Once the numerical
solution for the velocity and pressure fields in each cell of the
numerical grid is obtained, we compute the fluid velocity magnitudes $v$
associated to each channel. This value is the magnitude of the {\it
line average velocity vector} $\vec{v}$ calculated as the average over
the local velocity vectors ${\vec u}$ along the corresponding channel
width $l$.

In Fig.~3 we show the data collapse of all distributions of normalized
velocity magnitudes $P(v^{*})$, where $v^{*}=v/V$, obtained by
rescaling the variable $v^{*}$ with the corresponding value of
$\epsilon^{-2}$. It is also interesting to note that these rescaled
distributions follow a typical Gaussian behavior except for very small
$v^{*} \epsilon^{2}$, as indicated by the solid line in Fig.~3. In the
inset of Fig.~3 we also show that the average interstitial velocity indeed
scales with the porosity as $\langle v \rangle \sim \epsilon^{-2}$,
confirming the rescaling procedure adopted to obtain the collapse of
the distributions $P(v^{*})$ in the main plot of Fig.~3. Plotting for 
each channel $v$ against $l$ gives a cloud of points which for all
considered values of $\epsilon$ results in a rather unexpected least 
square fit relation of the type $v \sim \sqrt{l}$.
\begin{figure}
\includegraphics[width=8cm]{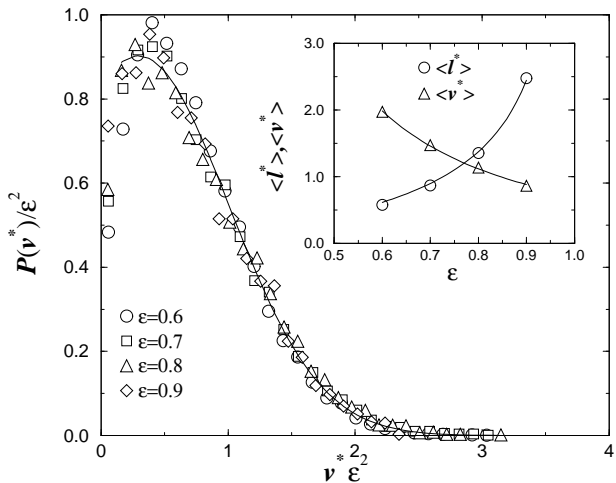}
\caption{Double-logarithmic plot of the distributions of the local 
normalized velocity magnitudes $v^{*}$, i.e., $v/V$, multiplied by
${\epsilon}^{2}$ as explained in the text. The solid line is a
Gaussian fit. The inset shows the dependence of $\langle l^{*}
\rangle$ and $\langle v^{*} \rangle$ on the porosity $\epsilon$.
The solid lines are the best fits to the data, corresponding  to 
$\langle l^{*} \rangle = a (\sqrt{b(1-\epsilon)}-1)$, with $a=1.22$ 
and $b=\pi/2\sqrt{3}$ (see Eq.~(\ref{Eq_Hans})) and $\langle v^{*} 
\rangle=0.71 \epsilon^{-2}$.}
\label{fig3}
\end{figure}

We now analyse the distribution of fluxes throughout the porous
medium. Each local flux $q$ crossing its corresponding pore opening
$l$ is given by $q=vl\cos\theta$, where $\theta$ is the angle between
$\vec{v}$ and the vector normal to the cross section of the channel
(see Fig.~1). In Fig.~4 we show that the distributions of normalized
fluxes $\phi=q/q_{t}$, where $q_{t}=V L_y$ is the total flux, have a
stretched exponential form, 
\begin{equation}
P(\phi) \sim \exp(-\sqrt{\phi/\phi_0})~,
\label{Eq_stretched}
\end{equation}
with $\phi_{0} \approx 0.005$ being a characteristic value. This
simple form of Eq.~(\ref{Eq_stretched}) is quite unexpected
considering the rather complex dependence of $P(l)$ on $\epsilon$.
Moreover, all flux distributions $P(\phi)$ collapse on top of each
other when rescaled by the corresponding value of ${\langle l^{*}
\rangle}^{-1} \epsilon^{2}$. This collapse for distinct porous media 
results from the fact that mass conservation is imposed at the
microscopic level of the geometrical model adopted here, which is
microscopically disordered, but at a larger scale is macroscopically
homogeneous \cite{Sahimi95}. As also shown in Fig.~4, it is possible to
reconstruct the distribution of fluxes using a convolution of the
distribution of velocities $v$ and the distribution of oriented
channel widths, namely $l {\rm cos}\theta$. Indeed, if we calculate
the integral,
\begin{equation}
P(\phi)\! = \!\int\! \int\! P(v) P(l \cos \theta) \delta(\phi-v l 
\cos \theta) dv d(l \cos \theta)~,
\label{Eq_convolution}
\end{equation}
we find that the original distribution $P(\phi)$ is approximately
retrieved, as can also be seen in Fig.~4 (solid line). 

Finally, the inset of Fig.~4 shows that the permeability of the
two-dimensional porous media closely follows the semi-empirical
Kozeny-Carman equation \cite{Dullien79},
\begin{equation}
\frac{K}{K_{0}}=\kappa \frac{\epsilon^{3}}{(1-\epsilon)^{2}}~,
\label{Eq_Kozeny-Carman}
\end{equation}
where $K_{0} \equiv h^{2}/12$ is a reference value taken as the
permeability of an empty channel between two walls separated by a
distance $h$. The proportionality constant $\kappa$ is given by
the following expression:
\begin{equation}
\kappa \equiv \left(\frac{D}{2h}\right)^{2} \frac{1}{\tau \alpha}~,
\label{Eq_kappa}
\end{equation}
where $\tau \equiv (L_{e}/L)^{2}$ is the {\it hydraulic tortuosity} of
the porous medium, $\alpha$ corresponds to the {\it pore shape factor}, 
and $L_{e}$ is an effective flow length \cite{Dullien79}. If we now make 
use of the Dupuit-Forchheimer assumption \cite{Dullien79},
\begin{equation}
\langle v \rangle=\frac{V}{\epsilon}\left(\frac{L_{e}}{L}\right)~,
\label{Eq_Dupuit-Forchheimer}
\end{equation}
we are led to the conclusion that the tortuosity of our porous medium
should also scale as $\tau \sim \epsilon^{-2}$. Considering the
validity of the Kozeny-Carman equation (\ref{Eq_Kozeny-Carman}) and
the definition of the constant $\kappa$ from Eq.~(\ref{Eq_kappa}), we
obtain as a consequence that the shape factor should behave as $\alpha
\sim \epsilon^{2}$.
\begin{center}
\begin{figure}
\includegraphics[width=8cm]{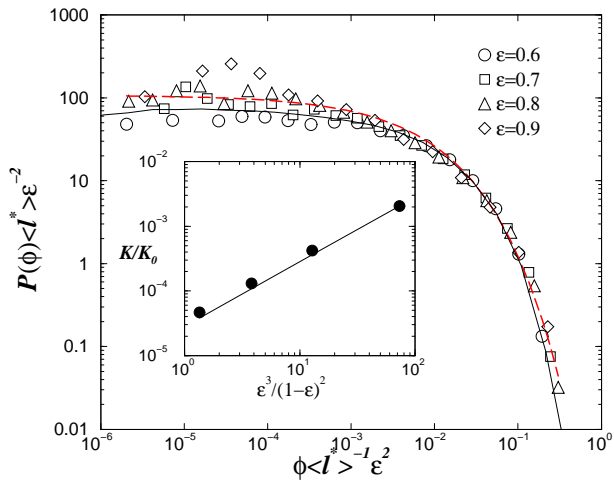}
\caption{Log-log plot of the distributions of the normalized 
local fluxes $\phi=q/q_{t}$ for different porosities $\epsilon$. The
(red) dashed line is a fit of the form $\exp(-\sqrt{\phi/\phi_{0}})$,
where $\phi_{0} \approx 0.005$. The full line stems from the
convolution as discussed in the text. In the inset we see a
double-logarithmic plot of the global flux and the straight line
verifies the Kozeny-Carman equation.}
\label{fig4}
\end{figure}
\end{center}
Summarizing we have found that although the distribution of channel
widths in a porous medium made by a two-dimensional RSA process is
rather complex and exhibits a crossover at $l \sim D$, the
distribution of fluxes through these channels shows an astonishingly
simple behavior, namely a square-root stretched exponential
distribution that scales in a simple way with the porosity. The
velocity magnitudes follow a Gaussian distribution truncated at small
velocities which scales with the square of the porosity. The
distribution of fluxes can be reconstructed as a convolution of the
velocity with the channel widths distributions corrected by the
velocity orientation factor $\cos\theta$. We propose simple scaling
laws for the local fluxes that deepen the understanding of the
intrinsic connection between geometrical and flow properties of the
random porous medium. Furthermore, we show that our results can be
macroscopically described in terms of the Kozeny-Carman
relation. Future tasks consist in generalizing these studies to
higher Reynolds numbers, three dimensional model of porous media and
other types of disorder. Other important challenges are to investigate
transient flow and tracer dynamics.

We thank Andr\'e Moreira, Salvatore Torquato and Bernard Derrida for
interesting discussions and the CNPq (CT-PETRO/CNPq), CAPES, FUNCAP,
FINEP and the Max Planck Prize for financial support.

\bibliographystyle{prsty}

\begin{thebibliography}{50}
\bibitem{Fluent}
{FLUENT} (trademark of {FLUENT} Inc.) is a commercial package for
computational fluid dynamics.
\end{thebibliography}

\end{document}